\documentclass[twocolumn,aps,prl,groupedaddress,showpacs,nofootinbib]
              {revtex4}%
\usepackage{graphicx}
\usepackage{amsmath}
\usepackage{bm}
\usepackage{relsize}
\RequirePackage{xspace}
\newcommand{\bqa}{\begin{eqnarray}}
\newcommand{\eqa}{\end{eqnarray}}
\newcommand{\pslash}{\slash\hspace{-0.55em}}
\newcommand{\as}{\alpha_{\mathrm{s}}}

\begin{document}


\title{\mbox{}\\[10pt]
Nonfactorizable $B\to\chi_{c0}K$ decay and QCD factorization}

\author{Ce Meng$~^{(a)}$, Ying-Jia Gao$~^{(a)}$, and Kuang-Ta Chao$~^{(b,a)}$}
\affiliation{ {\footnotesize (a)~Department of Physics, Peking
University,
 Beijing 100871, People's Republic of China}\\
{\footnotesize (b)~China Center of Advanced Science and Technology
(World Laboratory), Beijing 100080, People's Republic of China}}



\date{\today}

\begin{abstract}
We study the unexpectedly large rate for the
factorization-forbidden decay $B\to \chi_{c0}K$ within the QCD
factorization approach. We use a non-zero gluon mass to regularize
the infrared divergences in vertex corrections. The end-point
singularities arising from spectator corrections are regularized
and carefully estimated by the off-shellness of quarks. We find
that the contributions arising from the vertex and leading-twist
spectator corrections are numerically small, and the twist-3
spectator contribution with chiral enhancement and linear
end-point singularity becomes dominant. With reasonable choices
for the parameters, the branching ratio for $B\to\chi_{c0}K$ decay
is estimated to be in the range $(2-4)\times 10^{-4}$, which is
compatible with the Belle and BaBar data.
\end{abstract}


\maketitle


B meson exclusive decays to hadrons with charmonium are
interesting in studies of both strong interaction dynamics and
$CP$ violation. The naively factorizable decays \cite{BSW} such as
$B\rightarrow J/\psi K$ \cite{Cheng,Chay}, $B\rightarrow \eta_c K$
\cite{Chao02}, and $B\rightarrow \chi_{c1} K$ \cite{Chao03} were
studied in the QCD factorization approach \cite{BBNS} in which the
nonfactorizable vertex and spectator corrections were also
estimated.

To further explore the nonfactorizable contributions it is worth
studying the factorization-forbidden decays such as $B\rightarrow
\chi_{c0} K$. Recently, $B\rightarrow \chi_{c0} K$ decay has been
observed by Belle \cite{Belle02, Belle04} and BaBar \cite{BaBar04}
with surprisingly large branching ratio which is even comparable to
that of the factorization-allowed decay $B\rightarrow \chi_{c1} K$:
\begin{eqnarray}
 {\mathrm{Br}} (B^{+}\rightarrow \chi_{c0} K^+)&\!=\!&
(6.0^{+2.1}_{-1.8}\pm 1.1)\times10^{-4}~{\mbox[7]}, \nonumber\\
&=&(1.96\pm 0.35\pm 0.33)\times 10^{-4}~{\mbox[8]}, \nonumber\\
 {\mathrm{Br}} (B^{\pm}\rightarrow \chi_{c0} K^{\pm}) &=&
(2.7\pm0.7)\times10^{-4}~{\mbox[9]}.
\label{BeBar}
\end{eqnarray}
To explain the large decay rate of $B\rightarrow \chi_{c0} K$, the
final state re-scattering mechanism was suggested \cite{Cola}. On
the other hand, with the Light-Cone Sum Rules \cite{Melic} the
nonfactorizable contributions were found to be too small to
accommodate the observed $B\rightarrow \chi_{c0} K$ decay rate.

In fact, in Ref.~\cite{Chao03,Chao04} within the QCD factorization
approach it was found that for the $B\rightarrow \chi_{c0}K$ decay,
there exist both the infrared (IR) divergences in the vertex
corrections and the end-point singularities in the leading twist
spectator corrections. This implies that large nonfactorizable
contributions may come from soft gluon exchanges. As argued in
\cite{Chao04}, unlike the inclusive $B$ decays to charmonium, where
the IR divergences can be factorized into the color-octet matrix
elements associated with the higher Fock states of color-octet
$c\bar c$ with soft gluons \cite{Beneke99} in nonrelativistic QCD
(NRQCD) \cite{BBL1}, the IR divergences in the exclusive two-body
decays are difficult to be factorized.

At the qualitative level, the results of Ref.~\cite{Beneke99} may
suggest that some fraction of the large color-octet contribution
in the inclusive $B$ decays to charmonium does in fact end up in
two-body decay modes. That is, the soft gluon emitted by the
color-octet $c\bar c$ pair may be reabsorbed by the quarks in the
kaon or $B$ meson, leading to a possible connection between the
color-octet contribution and the infrared behavior of vertex and
spectator corrections in QCD factorization approach. So
qualitative estimates of these soft gluon contributions are
important for understanding the large branching ratios of both
$B\rightarrow \chi_{c0} K$ and $B\rightarrow J/\Psi K$.
Furthermore, since the $s$ quark emitted from the weak vertex
moves fast in the B meson rest frame, we may expect that the soft
gluon exchange is dominated by that between the $c\bar c$ pair and
the spectator quark.

In order to estimate the soft gluon contributions in these exclusive
decays, we may let the charm quark be off the mass shell or give the
gluon a mass, and then use the binding energy or gluon mass to
regulate the IR behavior. Introducing the binding energy
\cite{Barbieri} or momentum cutoff \cite{Novikov} was an useful way
in estimating the inclusive annihilation rates of $h_c$ and
$\chi_{c1}$ before the NRQCD factorization theory is developed. Even
after that, this approach is still widely used in quarkonium
phenomenology for qualitative estimates (see, e.g., \cite{Barnes}),
though it is not a rigorous theory. Recently, the binding energy
regularization was suggested in \cite{pham}. In the present paper,
we will estimate the non-factorizable decay rate for $B\rightarrow
\chi_{c0} K$ by using the gluon mass regularization for the IR
divergences, which is equivalent to the binding energy
regularization in physical nature. The main differences between our
calculations and those in Ref.~\cite{pham} are the treatments of the
end-point singularities in spectator interactions, which play the
most important role in numerical evaluations.

We treat charmonium as a color-singlet nonrelativistic (NR) $c\bar
c$ bound state.  Let $p$ be the total momentum of the charmonium and
$2q$ be the relative momentum between $c$ and $\bar c$ quarks, then
$v^2 \sim 4q^2/p^2 \sim 0.25$ can be treated as a small expansion
parameter \cite{BBL1}.  For P-wave charmonium $\chi_{c0}$, because
the wave function at the origin $\mathcal{R}_1(0)\!=\!0$, which
corresponds to the zeroth order in $q$, we must expand the amplitude
to first order in $q$. Thus we have (see, e.g., \cite{Kuhn})
\begin{eqnarray}
 \label{amp}
\mathcal{M}(B\to\!\! \chi_{c0}K)\!=\!\!\!\!\sum_{L_z,S_z}\!\langle
1L_z;1S_z|00\rangle
 \!\int\!\!\frac{\mathrm{{d}}^4 q}{(2 \pi)^3}q_\alpha\delta\!(q^0) \nonumber\\
 \times \psi_{1M}^\ast\!(q)
 \mathrm{Tr}[\mathcal{O}^\alpha\!(0)P_{1S_z}\!(p,\!0)
\!+\!\mathcal{O}\!(0)P^\alpha_{1S_z}\!(p,\!0)],
 \end{eqnarray}
where $\mathcal{O}(q)$ represents the rest of the decay matrix
element and can be further factorized as the product of $B \to K$
form factors and a hard kernel or as the convolution of a hard
kernel with light-cone wave functions of B and K mesons, within the
framework of QCD factorization approach. The spin-triplet projection
operator $P_{1S_z}(p,q)$ is constructed in terms of quark and
antiquark spinors as\footnote{To construct the spin-singlet
projection operator $P_{00}(p,q)$, one only needs  to replace
$\pslash \epsilon^\ast$ by $\gamma_5$ in Eq.(\ref{spinp}).}
\begin{eqnarray}
 \label{spinp}
P_{1S_z}(p,q)&\!=\!\!&\sqrt{\!\frac{3}{m_c}}\!\sum_{s_1,s_2}\!\!v(\frac{p}{2}\!-\!q,s_2)
 \bar u(\!\frac{p}{2}\!+\!q,s_1) \!\langle s_1;\!s_2|1S_z\!\rangle\nonumber\\
&\!\!=\!\!&-\sqrt{\frac{3}{4 M^3}}(\frac{\pslash p}{2}-\pslash
q-\frac{M}{2})\pslash \epsilon^\ast(S_z)(\pslash
p+M)\nonumber\\
 & &\times(\frac{\pslash p}{2}+\pslash q+\frac{M}{2}) ,
 \end{eqnarray}
and
 \bqa
\mathcal{O}^\alpha(0)\!&=&\!\frac{\partial
\mathcal{O}(q)}{\partial
q_\alpha}|_{q=0},\\
P^\alpha_{1S_z}(p,0)\!&=&\!\frac{\partial P_{1S_z}(p,q)}{\partial
q_\alpha} |_{q=0}. \eqa
In Eq. (\ref{spinp}) we take charmonium mass $M\simeq 2 m_c$ in NR
limit. Here $m_c$ is the charm quark mass.

The integral in Eq.~(\ref{amp}) is proportional to the derivative
of the P-wave wave function at the origin
 \bqa
\int\!\frac{\mathrm{{d}}^3 q}{(2 \pi)^3}q^\alpha \psi_{1M}^\ast(q)
=i\varepsilon^{\ast\alpha}(L_z)\sqrt{\frac{3}{4\pi}}
\mathcal{R}^{'}_1(0),
\eqa
 and we will use the following polarization relation  for
$\chi_{c0}$:
 \bqa\label{pol}
 \sum_{L_Z
S_Z}\!\!\!\varepsilon^{\ast\alpha}\!(\!L_z\!)\epsilon^{\ast\beta}\!(\!S_z\!)
\langle
1L_z;\!1S_z|00\rangle\!\!\!&=&\!\!\!\frac{1}{\sqrt{3}}(-\!g^{\alpha\beta}\!+\!\frac{p^\alpha
p^\beta}{M^2}\!).
 \eqa

Being contrary to $\chi_{c0}$, the K meson is described
relativistically by the light-cone distribution amplitudes
(LCDAs)~\cite{BBNS}:
\bqa
   \langle K(p\,')|\bar s_\beta(z_2)\,d_\alpha(z_1)|0\rangle&=&\nonumber\\
   &&\hspace*{-3.8cm} \frac{i f_K}{4} \int_0^1dxe^{i(y\,p\,'\cdot z_2+\bar y \,p\,'\cdot z_1)}
\Bigl\{ \pslash{p\,'}\,\gamma_5\,\phi_K(y)\nonumber\\
    &&\hspace*{-3.8cm}- \mu_K\gamma_5 ( \phi_K^p(y) - \sigma_{\mu\nu}\,p\,'^\mu (z_2-z_1)^\nu\,
    \frac{\phi_K^\sigma(y)}{6} ) \Bigr\}_{\alpha\beta},
\label{kaon}
  \eqa
where $y$ and $\bar{y}\!=\!1\!-\!y$ are the momentum fractions of
the $s$ and $\bar{d}$ quarks inside the K meson respectively, and
the chirally enhanced mass scale $\mu_K\! =\!
{m_K}^2/(m_s(\mu)\!+\!m_d(\mu))$ is comparable to $m_b$, which
ensures that the twist-3 spectator interactions are numerically
large, though they are suppressed by $1/m_b$ (see \cite{du}). The
twist-2 LCDA $\phi_K(y)$ and the twist-3 LCDA $\phi_K^p(y)$ and
$\phi_K^\sigma(y)$ are symmetric under
$y\!\leftrightarrow\!\bar{y}$ in the $SU(3)$ symmetry limit. In
practice, we choose the asymptotic forms for these LCDAs,
 \bqa \label{asy}
 \phi_K (y)=\phi_K^\sigma(y)=6y(1-y),\hspace{0.6cm}\phi_K^p (y)=1,
 \eqa

The effective Hamiltonian relevant for $B \to\! \chi_{c0}K$ is
\cite{BBL}
 \bqa
\mathcal{H}_{\mathrm{eff}}\!\!=\!\!\frac{G_F}{\sqrt{2}} \Bigl(\!
V_{cb} V_{cs}^*\!(C_1 {\cal O}_1\!+C_2 {\cal O}_2 )\!-V_{tb}
V_{ts}^* \sum_{i=3}^{6} C_i {\cal O}_i \!\Bigr),
 \eqa
where $G_F$ is the Fermi constant, $C_i$ are the Wilson
coefficients, and $V_{q_1q_2}$ are the CKM matrix elements. We do
not include the effects of the electroweak penguin operators since
they are numerically small. Here the relevant operators ${\cal
O}_i$ are given by
 \bqa
{\cal O}_1\!&=&\!(\overline{s}_{\alpha} b_{\beta})_{V-A} \cdot
(\overline{c}_{\beta} c_{\alpha})_{V-A},
 \nonumber\\
 {\cal O}_2\!&=&\!(\overline{s}_{\alpha} b_{\alpha})_{V-A} \cdot
(\overline{c}_{\beta} c_{\beta})_{V-A},
 \nonumber\\
{\cal O}_{3\!,\,5}\!&=&\!(\overline{s}_{\alpha} b_{\alpha})_{V-A}
\cdot \sum\nolimits_q (\overline{q}_{\beta} q_{\beta})_{V\mp A},
\nonumber\\
{\cal O}_{4\!,\,6}\!&=&\!(\overline{s}_{\alpha} b_{\beta})_{V-A}
\cdot \sum\nolimits_q (\overline{q}_{\beta} q_{\alpha})_{V\mp A},
\eqa
 where $\alpha,~\beta$ are color indices and the sum over $q$
runs over $u, d, s, c$ and $b$. Here $(\bar q_1 q_2)_{V\pm A}=\bar
q_1\gamma_\mu (1\pm\gamma_5) q_2 $.

 \begin{figure}[t]
\vspace{-3.0cm}
 \hspace*{-2.2cm}
\includegraphics[width=13cm,height=16cm]{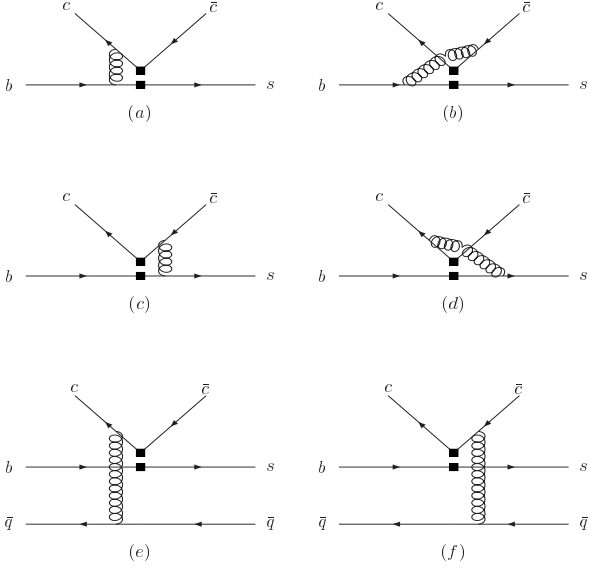}
\vspace{-5.0cm}
\caption{ Feynman diagrams for vertex and spectator corrections to
$B \to\! \chi_{c0} K$.} \label{fvs}
\end{figure}

According to \cite{BBNS} all non-factorizable corrections are due
to the diagrams in Fig.\ref{fvs}.  These corrections, with
operators ${\cal O}_i$ inserted, contribute to the amplitude
$\mathcal{O}(q)$ in Eq. (\ref{amp}), where the external lines of
charm and anti-charm quarks have been truncated. Taking
non-factorizable corrections in Fig.\ref{fvs} into account, the
decay amplitude for $B\to\! \chi_{c0}K$ in QCD factorization is
 \bqa\label{amp2}
  i\mathcal{M} =\frac{G_F}{\sqrt{2}}\Bigl[V_{cb}
V_{cs}^* C_1-V_{tb} V_{ts}^* (C_4 + C_6) \Bigr]\times A,
 \eqa
 and the coefficient
$A$ is given by
\bqa\label{a} A=\frac{i 6 \mathcal{R}^{'}_1(0) }{\sqrt{3\pi
M}}\frac{\alpha_s}{4\pi}\frac{C_F}{N_c} (1-z)F_1(M^2)
\frac{m_B^2}{M}\!\cdot\!\bigl( fI +   fI\!I \bigr).
 \eqa
Here $N_c$ is the number of colors, $C_F\!=\!(N_c^2-1)/(2 N_c)$,
and $F_{0,1}$ are the $B\! \to\! K$ form factors.  We have used
the relation ${F_0(M^2)}/{F_1 (M^2)}\!=\! 1\!-z\!$ \cite{Cheng,
Chay}, where $z\!=\!M^2/m_B^2\!\approx\! 4m_c^2/m_b^2$, to
simplify the amplitude in (\ref{a}). The function $fI$ is
calculated from the four vertex diagrams (a, b, c, d) in
Fig.\ref{fvs}, and $fI\!I$ is calculated from the two spectator
diagrams (e, f) in Fig.\ref{fvs}. The function $fI\!I$ receives
contributions from both twist-2 and twist-3 LCDAs of the K-meson,
and we can simply symbolize them as $fI\!I^2$ and $fI\!I^3$,
respectively.

In the statement of color transparency \cite{BBNS}, the IR
divergences should be cancelled between diagrams (a,b), (c,d) and
(e,f) respectively in Fig.\ref{fvs}. But it is not true when the
emitted meson is a P-wave charmonium, say, $\chi_{c0}$. The soft
gluons, which are emitted from a certain source, couple to the
charm and anti-charm quarks through both color charge and color
dipole interactions. For the color-singlet charm quark pair, the
total charge is zero, but the color dipole interactions, which are
proportional to the relative momentum $q$, give the leading order
contributions in both $1/m_b$ power expansion and NR expansion
(see Eq. (\ref{amp})). As a result, there exist IR divergences in
$fI$, while $fI\!I^{2,\, 3}$ suffer from logarithmic and linear
end-point singularities if the asymptotic form of kaon LCDAs are
used.

  Now we can express the function $fI$ as the Feynman parameter integrals
 \begin{equation} \label{f1}
 fI = -\int_0^1dx\int_0^{1\!-x}\!\!\!\!dy\;(d+c),
 \end{equation}
where the fuctions $d$ and $c$ are written by
\bqa\label{dc}
 d &=& \frac{2(1-z)}{(x+\frac y2 )(x+\frac
{z y}{2})} -\frac{(1-z)(1+z)xy}{((x+\frac y2 )(x+\frac {z
y}{2}))^2}\nonumber\\
&+&\!\!\frac{2(1-z)(1-x)}{\frac y2 ((z-1)x+\frac {z y}{2})}
+\frac{(1-z)^2xy(1-x)}{(\frac y2 ((z-1)x+\frac
{z y}{2}))^2} ~,\\
c&=&\frac{-\!y(z(3y\!+\!4)\!+\!2)\!+\!x(3y\!-\!z(9y\!+\!4)\!-\!2)}{(x+\frac
y2
)(x+\frac {zy}{2})}\nonumber\\
&+&\!\!\frac{\frac{1\!-\!z}{2}xy((2z\!-\!6)x^2\!+\!(6\!-\!zy\!-\!3y)x\!+\!y(z\!+\!2\!-\!zy))}
   {((x+\frac y2 )(x+\frac {zy}{2}))^2}\nonumber\\
&+&\!\!\frac{y(3x-2+z(3y-3x+2))}{\frac y2 ((z-1)x+\frac
{zy}{2})}\nonumber\\
&+&\!\!\frac{\frac{z-1}{2}xy^2(-3x+2+z(3x+y-3))}{(\frac y2
((z-1)x+\frac {zy}{2}))^2} ~.\nonumber \eqa

From Eq.~(\ref{dc}), we can see that the IR poles are all included
in the function $d$ before we integrate it completely. The first
two and the last two terms in $d$ come from diagrams (a,b) and
(c,d) in Fig.\ref{fvs}, respectively. The divergent integrals in
Eq. (\ref{f1}) can be regularized by non-zero gluon mass $m_g$, as
we have mentioned above. And the gluon mass pole in the divergent
integral is given by
 \bqa \label{div}
 \int_0^1dx\int_0^{1\!-x}\!\!\!\!dy\;d &=& \nonumber\\
 && \hspace{-2.8cm}-\frac{8
 z(1-z+\ln{z})}{(1-z)^2}\ln{(\frac{m_g^2}{m_b^2})}+\mbox{finite terms}.
 \eqa
Note that the infrared divergence would vanish if $z\to 0$ (i.e.,
if treating the charm quark as a light quark).

To derive functions $fI\!I^{2,3}$, we extract the light-cone
projector of $K$ meson in momentum space from Eq. (\ref{kaon}),
\bqa
   M_{\alpha \beta}^K(p\,')&=&
    \frac{i f_K}{4}\Bigl\{ \pslash{p\,'}\,\gamma_5\,\phi_K(y)\nonumber\\
    &&\hspace*{-1.5cm}- \mu_K\gamma_5 \bigl( \phi_K^p(y) - i\sigma_{\mu\nu}\,p\,'^\mu
    \frac{\partial}{\partial k_{2\nu}}
    \frac{\phi_K^\sigma(y)}{6} \bigr) \Bigr\}_{\alpha\beta}\,,
\label{Kprojector}
  \eqa
where $k_{2(1)}$ is the momentum of the anti-quark (quark) in $K$
meson, and the derivative acts on the hard-scattering amplitudes
in momentum space.

Using the projector given in~(\ref{Kprojector}) and eliminating
$\phi_K^{p,\sigma}(y)$ by (\ref{asy}), we get the explicit form of
$fI\!I^2$ and $fI\!I^3$:
 \begin{equation} \label{f22}
fI\!I^2 \!=\! a\! \int_0^1 d\xi \frac{\phi_B (\xi)}{\xi} \int_0^1
dy \frac{\phi_K (y)}{{\bar y}^2}[-2 z + (1 - z)\bar y ],
 \end{equation}
 \begin{equation} \label{f23}
 fI\!I^3 \!\!=\!\! \frac{a\cdot r_K}{1-z}\! \int_0^1 d\xi \frac{\phi_B
(\xi)}{\xi}\! \int_0^1 dy \frac{1}{{\bar y}^2}[3 z - (1 - z)\bar y
],
 \end{equation}
where the factor $a$ is defined as
\begin{equation}\label{afactor}
 a=\frac{8\pi^2 f_K f_B}{N_c(1-z)^2 m_B^2F_1(M^2)}~.
 \end{equation}
Here, due to our definition of $fI\!I$ in (\ref{a}), the form factor
$F_1$ is present in the denominator.

In Eqs.~(\ref{f22}, \ref{f23}), $\xi$ is the momentum fraction of
the spectator quark in the $B$ meson and
\begin{equation}\label{rk}
r_K(\mu) = 2m_K^2/[{m_b(\mu)(m_s(\mu)+m_d(\mu))}]
\end{equation}
is of order one and can not be neglected. The integral over $\xi$
is conventionally parameterized as \cite{BBNS}
\begin{equation}
 \int_0^1 d\xi \frac{\phi_B(\xi)}{\xi}=\frac{m_B}{\lambda_B}~.
 \end{equation}

If we choose the asymptotic form of the LCDAs of kaon in
Eq.~(\ref{asy}), we will find logarithmic (linear) singularities
in $fI\!I^2$ ($fI\!I^3$). As we have mentioned before, these
singularities came from the color dipole interactions between the
soft gluons and the P-wave charm quark pair, just like what
happened in the vertex corrections.

Our result of $fI\!I^2$ is consistent with the previous ones
\cite{Chao04,pham}. However, our function $fI\!I^3$ is different
from that in Ref. \cite{pham} because they used a twist-3 light-cone
projector
\begin{equation}\label{kpro2}
M_{\alpha\beta}^K=\frac{if_K}{4}\Bigl\{\pslash{p\,'}\,\gamma_5\,\phi_K(y)
    - \mu_K\gamma_5 \frac{\pslash{k_2}\pslash{k_1}}{k_2\cdot k_1} \phi_K^p(y)\Bigr\}_{\alpha\beta} \,,
\end{equation}
which could be derived from Eq.~(\ref{Kprojector}) by adopting an
integration by parts on $y$ and dropping the boundary terms. If
one simply parameterizes the linear singularities in
Eq.~(\ref{f23}) as
\begin{equation}\label{endpoint}
\int\frac{dy}{y^2}=\frac{m_B}{\Lambda_h}~,
\end{equation}
where $\Lambda_h\!\sim\!500~\!\mbox{MeV}$ is the infrared cutoff,
we will find a large difference between our results and those in
Ref.~\cite{pham}. Since the decay rate is very sensitive to the
numerical value of $fI\!I^3$ as one can see in the following, we
should be very careful in regularizing these singularities.

The key point here is that the boundary terms have the form
$\frac{\phi_K^\sigma(y)}{y^2}|^1_0$ and can not be dropped safely,
when the asymptotic form of $\phi_K^\sigma(y)=6y\bar{y}$ is
inserted. So the integration by parts is not well-defined when one
simply parameterizes the linear singularities as in
Eq.~(\ref{endpoint}) and in Ref.~\cite{pham}. That is why our
result of $fI\!I^3$ is different from that in Ref.~\cite{pham}.

The above problem comes from the observation that the product of
virtualities of the propagators in Fig.1 (e,f) goes to zero faster
than $\phi_K^\sigma(y)$ in the end point regions. However, if we
regularize all these small-virtualities carefully (i.e., introducing
small off-shellness or transverse momenta for quarks and gluons), as
suggested in Ref.~\cite{Beneke02}, we can see that the boundary
terms are exactly zero,
\begin{equation}\label{boundary}
\frac{y^n\phi_K^\sigma(y)}{(y+\lambda)^{n+2}}|^1_0=0,\hspace{0.6cm}
n=0,1,2~...~,
\end{equation}
and then the integration by parts is well-defined. That is, the two
projectors are equivalent if and only if one regularizes the linear
singularities properly, e.g.,
\bqa\label{endpoint2} \int \frac{dy}{y^2} \rightarrow \int
\frac{dy}{(y+\lambda)^2} = \frac{1}{\lambda}-1+O(\lambda)~,\nonumber \\
 \int \frac{y dy}{y^3}\rightarrow \int \frac{y
dy}{(y+\lambda)^3} = \frac{1}{2\lambda}-1+O(\lambda)~...~, \eqa
where the relative off-shellness $\lambda$ should be of order of
$\Lambda_h/m_B$. In practice, one needs to be careful and note that
the integral kernels $1/y^2$ and $y/y^3$ give different
contributions in the scheme described by Eq.~(\ref{endpoint2}),
although superficially they are the same in the expression of
Eq.~(\ref{f23}) and in the parametrization of Eq.~(\ref{endpoint}).

It is worth emphasizing that this scheme is more physical than the
one in Eq.~(\ref{endpoint}) since the off-shellness or transverse
momenta of quarks and gluons are naturally serve as infrared cutoffs
when $y\rightarrow 0$. Furthermore, it is a proper scheme to realize
the factorization for the electromagnetic form factors of $\pi$ at
twist-3 level~\cite{Beneke02}.

To determine the value of $\lambda$, we introduce the binding energy
$b\!\simeq\! M-2m_c\!>\!0$ \cite{pham} for $\chi_{c0}$, and then the
virtuality of the quark line in Fig.1(e) or (f) goes to
\begin{equation}\label{virtuality}
(p_c+\bar{y}p')^2-m_c^2\simeq
\frac{(1-z)}{2}m_B^2(\bar{y}+\frac{z}{1-z}\frac{b}{M}\,).
\end{equation}
Here it is evident that the last term in the last parentheses plays
the role of $\lambda$. In the APPENDIX we will show that the
relative off-shellness determined by the gluon propagator is of the
same sign and the same order as $\lambda$. In practice, we use
\begin{equation}\label{lambda}
\lambda=\frac{z}{1-z}\frac{b}{M}\,.
\end{equation}

Following the scheme in Eq.~(\ref{endpoint2}) and using the
asymptotic forms of LCDAs in  Eq.~(\ref{asy}), we reexpress the
function $fI\!I^3$ as
\bqa \label{f23explicit} fI\!I^3
\!\!\!&=&\!\!\!\frac{a\!\cdot\!r_K}{1-z}\!\cdot\!\frac{m_B}{\Lambda_B}
\! \int_0^1\! dy[\frac{3z}{(y+\lambda)^2}-\frac{3zy+(1-4z)y^2}{(y+\lambda)^3}\nonumber\\
&&+\frac{3zy^2-3zy^3}{(y+\lambda)^4}].
 \eqa
For simplicity, here we have used the same relative off-shellness
$\lambda$ to regularize each factor $y$ in denominators. If we set
$\lambda=0$ in  Eq.~(\ref{f23explicit}), this expression will fall
back upon the form given in  Eq.~(\ref{f23}). We also derive the
function by using the projector in  Eq.~(\ref{kpro2}), and get a
different expression,
\bqa\label{f23zhu}
fI\!I^3\!\!\!=\!\!\!\frac{a\!\cdot\!r_K}{1-z}\!\cdot\!\frac{m_B}{\Lambda_B}
\! \int_0^1\! dy[\frac{3z}{(y+\lambda)^2}\!-\!\frac{z
y+(1-z)y^2}{(y+\lambda)^3}]. \eqa
Similarly, up to a universal normalization factor, the function
$fI\!I^3$ will fall back on the same form given in Ref. \cite{pham}
if $\lambda$ is set to be zero. Completing the integrals in
 Eq.~(\ref{f23explicit}) and  Eq.~(\ref{f23zhu}), we get the same result
\bqa \label{f23final}
fI\!I^3\!\!\!=\!\!\frac{a\!\cdot\!r_K}{1-z}\frac{m_B}{\Lambda_B}[\frac{5z}{2\lambda}
\! +
\!(1\!-\!z)ln\lambda\!+\!\!\frac{1}{2}(3\!-\!7z)\!\!+\!O(\lambda)],
\eqa
as it should be. Again, we should emphasize that the emergence of
infrared divergences and end-point singularities in decay
amplitudes partly destroys the factorization assumption, and the
soft interaction mechanism may be dominant in this decay mode.
Before the divergences are removed or absorbed in some other
factorization schemes, Eq.~(\ref{f23final}) can reasonably serve
as a model dependent estimation for soft gluon effects
contributing to twist-3 spectator interactions. Furthermore,
because of the linear singularities in  Eq.~(\ref{f23final}),
$fI\!I^3$ is not power suppressed in $1/m_B$, rather it is
chirally and kinematically enhanced.

To be consistent with $fI\!I^3$, we regularize $fI\!I^2$ in the
 same scheme and get
\bqa \label{f22final}
fI\!I^2\!\!=\!\!a\!\cdot\!\frac{m_B}{\Lambda_B}[12zln\lambda+21z+O(\lambda)].
 \eqa

For numerical estimates, we use the following input parameters
with two values for the QCD scale $\mu\!=\!\sqrt{m_b
\Lambda_h}\!\approx\! 1.45$ Gev and $\mu\!=\! m_b\! \approx\! 4.4$
Gev:
 \bqa \label{parameter}
&&\!\!  M
=3.42 ~\mbox{GeV}, ~ \!\! m_B=5.28~\mbox{GeV},~\!\! \lambda_B\!=\!300~\!\mbox{MeV}, \nonumber \\
&& \!\!f_B\!=\!216 ~\mbox{MeV}\mbox{\cite{Gray}}, ~
\!\!f_K\!=\!160~\mbox{MeV},
~\!\!F_1(M^2)=0.75~\mbox{\cite{Ball}},\nonumber \\
&&\!\!\mathcal{R}^{'}_1(0)\!=\!\sqrt{0.075}~\!\mbox{GeV}^{5/2}\mbox{\cite{Quig}},~C_1(\mu)=1.239(1.114),\nonumber\\
&&\!\!C_4(\mu)=-0.046(-0.027),~
C_6(\mu)=-0.068(-0.033),\nonumber\\
&& \!\!\as(\mu)=0.34(0.22),~r_K(\mu) =0.85(1.3),~\lambda=0.087.
 \eqa
In  Eq.~(\ref{parameter}) the $\mu$-dependent quantities at
$\mu\!=\!1.45$ Gev ($\mu\!=\!4.4$ Gev) are shown without (with)
parentheses. The Wilson coefficients $C_i$ are evaluated at
leading order by renormalization group analysis \cite{BBL}, since
the amplitudes in  Eq.~(\ref{a}) are of leading order in $\as$.

The numerical results of $fI$  are listed in Tab.\ref{f} with the
gluon mass varying from 200 Mev to 500 Mev (the typical scale of
$\Lambda_{QCD}$). Comparing $fI$ with the values of
$fI\!I^{2,\,3}$, we find that for $B\rightarrow \chi_{c0}K$, both
$fI$ and $fI\!I^2$ are small and they are partially canceled. As a
consequence, the prediction for Br($B\rightarrow \chi_{c0}K$)
would be about an order of magnitude smaller than
Eq.~(\ref{BeBar}) if only the leading-twist functions were used
(see Tab.\ref{br}). However, the chirally enhanced twist-3
contribution is numerically large and makes the predicted decay
rate to be comparable to the experimental data
 Eq.~(\ref{BeBar}).

\begin{table}[t]
\caption{Functions evaluated by using the parameters in
 Eq.~(\ref{parameter}). The $\mu$-dependent values given at
$\mu\!=\!1.45$ Gev ($\mu\!=\!4.4$ Gev) are shown without (with)
parentheses.}
\label{f}\begin{tabular}
 {c|c|c|c}\hline
                 & $fI$           & $fI\!I^2$      & $ fI\!I^3(\mu)$ \\\hline
  $m_g=0.5$ Gev  & ~12.9-17.7$i$~ & ~~~~$-$7.8~~~~ & ~~~~35.0(53.6)~~~~ \\
  $m_g=0.2$ Gev  & ~20.7-19.3$i$~ & ~~~~$-$7.8~~~~ & ~~~~35.0(53.6)~~~~ \\
   \hline
\end{tabular}
\end{table}
\begin{table}[t]
\caption{Theoretical predictions for the branching ratio of
$B\!\!\rightarrow \!\!\chi_{c0}K$ including twist-3 contributions.
For comparison, results without twist-3 contributions are listed in
the parentheses.}
\label{br}\begin{tabular}{c|c|c}\hline
   $10^5\times Br$  & $m_g=0.5$ Gev & $m_g=0.2$ Gev \\\hline
 $\mu\!=\!1.45$ Gev     &  30~(5.4)     &  42~(8.6)    \\
 $\mu\!=\!4.40$ Gev     &  19~(1.7)     &  24~(2.7)    \\\hline
\end{tabular}
\end{table}

Our predictions  for the branching ratio of $B\rightarrow
\chi_{c0}K$  are listed in Tab.\ref{br}, and the values in the
parentheses are the results evaluated by using leading-twist
contributions only. From Tab.\ref{br} we can see that the results
are not very sensitive to the value of the gluon mass. For
comparison, we follow the available results in Ref.~\cite{pham},
where the IR divergences are regularized by binding energy, to
evaluate $fI$ and give Br$(B\!\!\rightarrow
\!\!\chi_{c0}K)\!\!=\!\!22(16)\!\!\times \!\!10^{-5}$ for
$\mu\!\!=\!\!1.45(4.4)$ Gev. Again, the conclusion is similar to
the gluon mass scheme. We see that by treating the singularities
in spectator interactions properly, we can obtain the large
experimental rate of $B\!\!\rightarrow \!\!\chi_{c0}K$ without
introducing any unknown imaginary part for spectator interactions
\cite{pham}.

In summary, We have studied the factorization-forbidden decay
$B\!\to\!\chi_{c0}K$ within the framework of the QCD factorization
approach. We use the gluon mass to regularize the infrared
divergence in vertex corrections. The end-point singularities
arising from spectator interactions are regularized and estimated
carefully by the off-shellness of quarks in the small virtuality
regions. We find that for this decay the contributions from the
vertex and leading-twist spectator corrections are numerically
small, and the twist-3 spectator contribution with chiral
enhancement and linear end-point singularity becomes dominant.
With reasonable choices for the parameters, the
$B\!\!\to\!\!\chi_{c0}K$ decay branching ratio is estimated to be
in the range of $(2\!-\!4)\!\times \!10^{-4}$ , which is
compatible with the Belle and BaBar data. We would like to point
out that there are other B exclusive decays to charmonia such as
$B\!\!\to\!\!\chi_{c2}K$, $B\!\!\to\!\!h_cK$ as well as
$B\!\!\to\!\!\psi(3770)K$, and it is worth studying those decays
to see if the soft spectator interactions are also the dominant
mechanisms\cite{Meng06}.

$Note.~$ Since this result was reported in arXiv:hep-ph/0502240, the
$B\!\!\to\!\!\chi_{c0}K$ decay has also been studied with
$k_T$-factorization in the PQCD approach \cite{Li}, in which the
vertex corrections are ignored and the spectator corrections are
found to give a large enough decay rate in comparison with
experiment. Their result is consistent with ours in the sense that
in both approaches the vertex corrections are found to be small, and
the spectator corrections give dominant contributions to the decay
rate. For the vertex corrections, the authors of Ref.~\cite{Li} take
them into account through the variation of renormalization scale for
the factorizable contributions, while we use the gluon mass (or the
binding energy) to regularize the IR divergence and qualitatively
estimate the size of these corrections. For the spectator
corrections, in our approach the dominant contribution comes from
soft gluon exchange or quark off-shellness, which are related to the
end point singularities when the transverse momentum in the kaon is
ignored. It is therefore interesting to examine if the main
contributions in Ref.~\cite{Li} are also from the regions with small
virtualities for quarks.

\begin{acknowledgments}
ACKNOWLEDGMENTS

We thank H.Y. Cheng for brining Ref.\cite{Belle04} to our
attention and for helpful comments. This work was supported in
part by the National Natural Science Foundation of China (No
10421003) and the Key Grant Project of Chinese Ministry of
Education (No 305001).
\end{acknowledgments}

\appendix
\section{Appendix}
In this APPENDIX we will show that the relative off-shellness
determined by the gluon propagator in Fig1.(e) or (f) is of the same
order as $\lambda$ in (\ref{lambda}). To be explicit, we use the
transverse momentum of the spectator quark to regularize the gluon
propagator just like that in PQCD approach \cite{Li2001}.

Let $l$ denote the momentum of the spectator quark in B meson, then
the denominator in the gluon propagator is $(l-k_2)^2\simeq
-(1-z)\xi\bar{y}m_B^2$ if the transverse component $\vec{l}_{\bot}$
and $\vec{k}_{\bot}$ are ignored. Then the transverse momenta can be
used to regularize the gluon propagator near the end-point region
$\bar{y}\rightarrow0$, and the denominator in this non-zero
transverse momentum scheme can be written as \cite{Li2001}
 \bqa \label{gluon virtuality}
 (l-k_2)^2&\simeq&-(1-z)\xi\bar{y}m_B^2-(\vec{l}_{\bot}-\vec{k}_{\bot})^2\nonumber\\
 &=&-(1-z)m_B^2(\bar{y}+\frac{(\vec{l}_{\bot}-\vec{k}_{\bot})^2}{(1-z)\xi
 m_B^2}).
 \eqa
 Obviously, here the average value $\langle\frac{(\vec{l}_{\bot}-\vec{k}_{\bot})^2}{(1-z)\xi
 m_B^2}\rangle$ serves as the relative off-shellness $\lambda^{'}$ for
 the virtual gluon. Note that all the components of $l$ should be
 of order $\bar{\Lambda}=m_B-m_b$, it is then easy to determine that $\lambda^{'}\sim
 O(\bar{\Lambda}/m_B)$, which is of the same order as $\lambda$ in
 ($\ref{lambda}$). Furthermore, both $\lambda$ and
 $\lambda^{'}$ are of positive values, hence $fI\!I^{(2,3)}$ in Tab.I do not contain imaginary
 parts.


\end{document}